# Tactile Melodies: A Desk-Mounted Haptics for Perceiving Musical Experiences

Moora Raj Varshith[1] and Gowdham Prabhakar[2]

[1] Design Department, Indian Institute of Technology Kanpur, Kanpur, Uttar Pradesh, 208016, India
[2] Design Department, Indian Institute of Technology Kanpur, Kanpur, Uttar Pradesh, 208016, India
`varshith@iitk.ac.in`

**Abstract.** This paper introduces a novel interface for experiencing music through haptic impulses to the palm of the hand. It presents a practical implementation of the system exploring the realm of musical haptics through the translation of MIDI data from a Digital Audio Workstation (DAW) into tactile sensations, from a set of haptic actuators, in real-time. It also includes a suitable music-to-haptic mapping strategy to translate notes from musical instruments to haptic feedback. The haptic actuators, placed strategically on the palmar surface of the hand allowed users to perceive music and were able to identify melody and rhythm of different musical compositions. A pilot user study conducted intended to assess the accuracy of the interface by testing the participants to select the correct audio presentation from the haptic presentation of the same musical composition. It presents a comparative study, differentiating between those with prior musical background and those without, in identifying the correct auditory counterpart solely through haptic inputs. This pilot study delves into how users perceive and interpret haptic feedback within the context of musical compositions. The pilot study showed promising results in enriching our understanding of user responses to haptic feedback in musical scenarios and exploring the intricacies of user experience with the system and its impact on musical interpretation.

**Keywords:** Haptics, Haptic Interface, DAW, MIDI, Tactile, audio-tactile mapping.

## 1    Introduction

### 1.1    Haptics in Music Technology

Research in haptic feedback has extended into music technology, aiming to enhance musicians' experiences and provide sensory substitution systems for the hearing impaired. Haptic Music Players (HMPs) enable users to experience music through touch by using vibrotactile feedback to translate musical elements into tactile sensations. This makes music accessible to those with hearing impairments and enhances the overall sensory experience for all users. HMPs often use small, lightweight actuators like eccentric rotating mass (ERM) vibration motors to generate tactile feedback, chosen for their portability and ability to provide perceptible vibrations within a specific frequency range. The challenge in designing HMPs lies in accurately translating musical elements such as melody, pitch, rhythm, and timbre into tactile feedback. This paper aims to provide a prototype for a desk-mounted haptic music player that can be accessed through the user's hand and provide a suitable mapping strategy to convey music to the hand through haptic actuators.

## 1.2  Haptic Mapping Strategies

Various mapping strategies have been put forward by researchers working in the field of musical haptics depending on the mode of interface and type of music they were aiming at. One approach to mapping audio music to haptics is through direct frequency mapping, where the frequency of musical notes is mapped directly to the frequency of the haptic actuators, allowing users to feel the pitch of the notes through variations in vibration intensity. Another approach is spatial arrangement, where actuators are spatially arranged on the skin, with each actuator assigned to a specific band of frequency. This paper investigates the second approach, calibrating the intensity and purpose of the haptic actuator to the distribution of touch receptors of the palmar surface of the hand. This frees up the intensity parameter to convey other musical attributes like timbre.

## 1.3  User Study on Haptic Mapping Interfaces

A pilot study was conducted to evaluate the effectiveness of translating music into haptics and understand user perceptions of haptic feedback in musical contexts. Participants interacted with a haptic feedback system designed to translate musical snippets into tactile sensations, using ERM motors controlled by an Arduino with MIDI signals guiding the haptic output. After familiarizing themselves with the melodies through auditory exposure, participants identified the songs based on haptic feedback alone. The study revealed that participants, especially those with musical training, could recognize and differentiate between musical pieces based on haptic feedback. This underscores the potential of haptic interfaces to convey complex musical information and enhance sensory experiences in innovative ways.

## 1.4  Use Cases and Future Scope

Desk-mounted haptic music players offer a versatile solution with the potential to revolutionize musical experiences by providing immersive tactile feedback synchronized with audio, enhancing engagement in virtual reality concerts and making music accessible to the hearing impaired. These devices also hold promise in music education, facilitating faster learning of variety of percussion instruments such as piano, tabla, etc., through additional sensory feed-back, aiding in proper technique and finger positioning. Furthermore, haptic music players can extend their utility to meditation and relaxation practices, enhancing sensory experiences for deeper relaxation and focus [1]. With a working system in place, interested researchers can come up with more creative ways of utilizing this technology and conduct long-term studies about the benefits of integrating haptics in our lives.

The following Section presents the relevant existing literature on the field of musical haptics and audio tactile rendering mechanisms and how they were adopted in haptic music players. Section 3 deals with how this data formulated the design of our desk-mounted haptic player. Section 4 explains audio-to-haptic mapping system that was employed for this prototype and the reasoning behind it. Section 5 presents the pilot user study that was conducted followed by its results and conclusion. Lastly, Section 8 offers the potential use cases and future studies to be conducted in the field of this technology.

## 2    Background

The book "Musical Haptics" by S Papetti provides a foundational overview of this field, discussing the tactile aspects of music performance, the role of haptic feedback in digital musical interfaces, and the impact on user experience and performance [2]. The paper "Audio-Tactile Rendering: A Review on Technology and Methods to Convey Musical Information through the Sense of Touch" by Remache-Vinueza, B. et al. provides a detailed collection explores the technology and methods used to convey musical information through touch, offering insights into audio-tactile interactions [3]. It introduces a method of using musical metaphors as a language to translate audio music with haptic music and offers examples of different strategies implemented.

In plenty of user studies conducted on wearable haptic music players such as in 'Feel Music'[4], 'Mood Glove'[5] and 'Haptic Hand'[6] , results have shown integrating haptics with audio has proven to enhance the experience by accentuating the musical elements, providing a more emotional communication with the medium. These studies also provided a system to conduct user studies of haptic music players using Hevner's circle to get an understanding of the affective qualities of vibrotactile sensations [4]. We also learn about the methodology of transmitting the music information from MIDI files to the actuators using microcontrollers, which is the general standard in building a haptic music player currently.

In the process of building the prototype, we employed Digital Audio Workstations (DAWs) which are software applications essential for recording, editing, and producing audio and MIDI data on computers, offering versatile tools for music production and sound design. DAWs excel in handling MIDI information, allowing precise recording, editing, and transmission to electronic devices. Their features include MIDI sequencing for note and controller data, virtual instruments for custom sounds, MIDI mapping for control, flexible routing to external devices, and MIDI effects for creative manipulation. This versatility enables seamless integration with custom electronic devices, facilitating real-time control and innovative musical experiences through the transmission of MIDI data. This real-time communication with the haptic music player through the computer is one of the key challenges we had to solve which would let us test and iterate our audio to haptic mapping strategies.

Most of the musical haptic interfaces are used for instrument training and feedback for musicians, but not much for musical experience [7]. Studies have explored the use of haptic feedback in music-focused exercises, presenting findings that analyze its impact within a musical context [7]. Investigations have been conducted to design meaningful ways to convey musical information via the sense of touch, emphasizing the importance of transparent and immersive musical experiences facilitated by vibrotactile feedback and stimulation [7]. The vibrotactile feedback is mostly generated using small and lightweight eccentric rotating mass (ERM) vibration motors, which are suitable for enhancing the portability of the haptic interface [8]. However, independent control of amplitude and frequency of vibrations is not possible, and mapping of signals carrying temporal components of music such as rhythm to ERMs may not be accurate [8]. The research highlights the need for further advancements in the field of musical haptics to enhance the overall user experience and make haptic music technology more accessible and efficient [7][8].

## 3    Design

We developed a tactile feedback system mounted on a desk, accessible by the right hand. This system incorporates 10 Eccentric Rotating Mass (ERM) Coin Vibrators positioned to stimulate the fingertips of the fingers. The activation and intensity control of these vibrators are managed by an Arduino Mega 2560 Rev3. Initially, our approach involved processing audio input and mapping frequency ranges onto the motors, aiming to convey the melody of music. However, the frequency analysis of polyphonic tones did not result in accurate output as the algorithms involving music information retrieval are computationally difficult and still are being studied. [9]

### 3.1    Distribution of touch receptors of the palmar region of hand

Touch receptors in the palm of our hand play a crucial role in tactile sensation and perception. These receptors, including Merkel's discs, Meissner's corpuscles, Pacinian corpuscles, and Ruffini's endings, are specialized mechanoreceptors that respond to diverse types of stimuli. Merkel's discs, densely located in the fingertips, are ideal for processing information about shape and texture. Meissner's corpuscles, with small receptive fields, excel in transmitting details about movement and texture, aiding in grip maintenance. Pacinian corpuscles, with large receptive fields, are effective in detecting vibrations caused by objects in contact with the hand, crucial for tool use. Ruffini's endings, slowly adapting receptors, primarily respond to skin stretching, contributing to awareness of finger and hand position. These receptors collectively contribute to our ability to perceive and interact with the tactile world, enabling functions like object recognition, texture discrimination, sensory-motor feedback, and social exchange [10][11].

Our device would primarily engage with Pacinian corpuscles and hence the distribution density of these receptors on the palmar region played a crucial role in the placement of haptic actuators. According to a study by Stark B et al., [12] the fingers contain the majority of the corpuscles in the hand, with a total mean number of 300. Specifically, 44 to 60% of these corpuscles are located in the fingers. Which make them the most sensitive to vibrations. The size of Pacinian corpuscles are smaller in the fingers compared to the palmar region giving them a higher resolution in distinguishing the frequency of vibrations. 23 to 48% in the metacarpophalangeal area, and 8 to 18% in the thenar and hypothenar regions. Notably, corpuscles found in the palmar skin over the distal phalanx are smaller than those in the metacarpophalangeal area, rendering their sensitivity and low tactile spatial resolution. Additionally, the lowest density of corpuscles is found along the nerves and vessels of the middle phalanx. The threshold of human perception of vibration starts from 0.3Hz to 1000Hz [7]. The ERM motors we employed have a perceptible frequency of range of 50Hz to 150Hz.

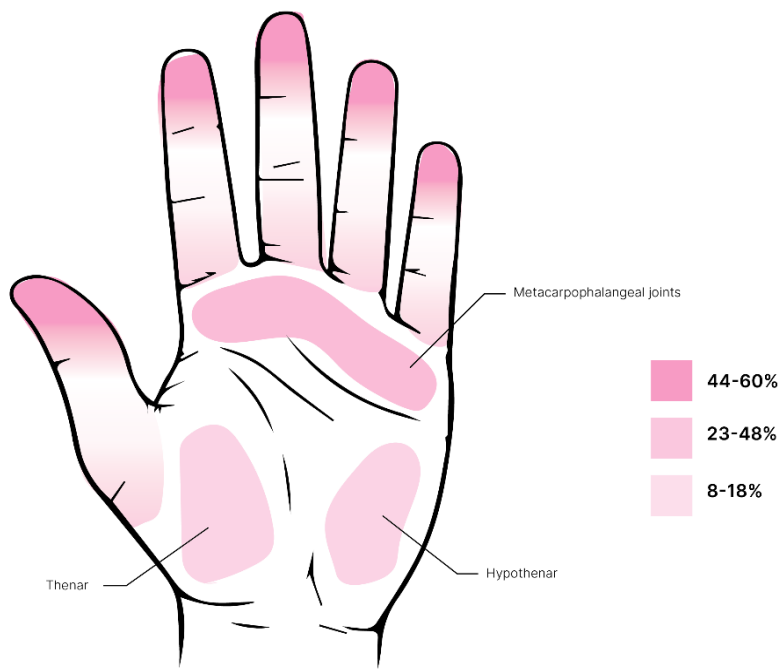

**Fig. 1.** Distribution of Human Pacinian Corpuscles in the Hand

### 3.2   Actuator Placement

Taking note of the distribution of Pacinian corpuscles, and the area of contact of the palmar region upon resting on flat surface the placement of actuators is decided. Five actuators are placed under all the fingertips, owing to their maximum sensitivity and high resolution of receptors. Three motors are placed under the metacarpophalangeal joints and one actuator is given to the thenar region and one actuator to the hypothenar region.

The dimensions of the prototype are taken from the mean hand measurements given in "Indian Anthropometric Dimensions for Ergonomic Design Practice" by Chakrabarti, D [13]. The Hand length, the distance from the base of the palm to the tip of the middle finger is 178 mm. the palm length, which is the distance from the base of the palm to the base of the middle finger is 103 mm. The hand breadth, without the thumb at the metacarpal is 80 mm. Moreover, the fringes are given around the finger to propagate the vibrations throughout the finger, and to accommodate for the variation in the length of fingers of different users.

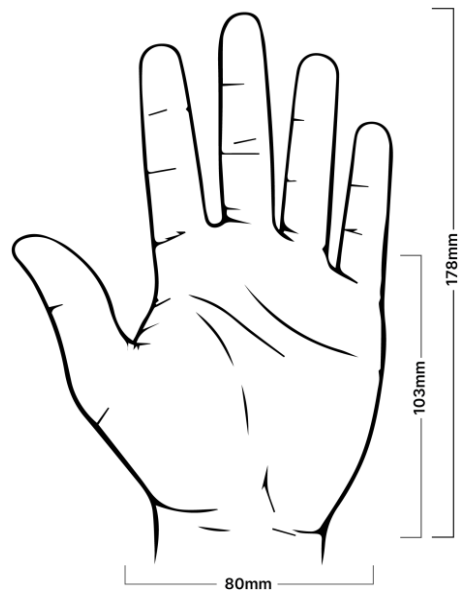

**Fig. 2.** Dimensions of Hand Length, Palm Length and Hand Breadth of average Indian human hand

The Prototype is made from acrylic board using laser-cutting to cut the fringes and holes for the ERM motors. A slight curvature is given to the board to increase the surface of contact of the thumb, providing an easier geometry to grab on. The ERM motors are attached through the holes from behind, which are then connected to the Arduino Mega through jumper wires. A heavy wooden plank is used as base support to dilute the vibrations of the motors onto the table and reduce wastage of vibrations.

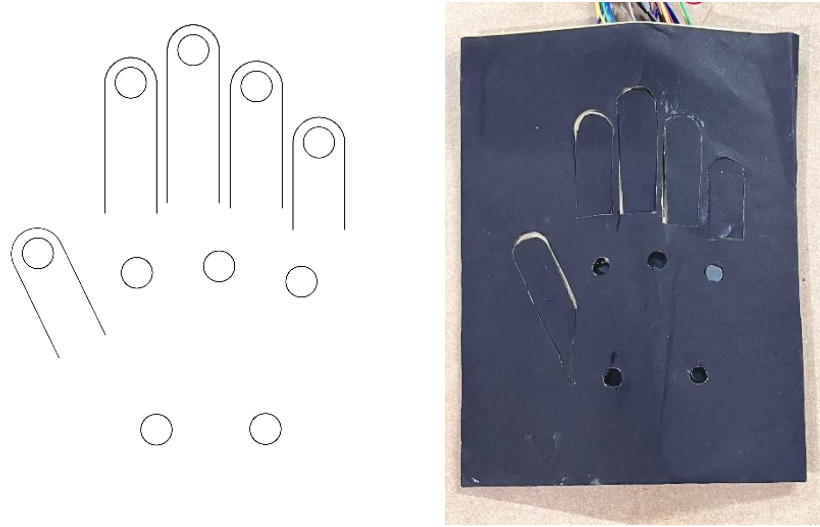

**Fig. 3.** CAD model of the desk mounted haptic music player casing (left) and Final Prototype of the same (right)

### 3.3 Utilizing MIDI protocol for Signal Transmission to Arduino

Musical Instrument Digital Interface (MIDI) protocol is used by all electronic instruments for seamless real-time communication between instrument and computer. It is used to play and record performances on any electronic instrument accessible on a computer. It does not transmit any audio information but rather musical information regarding notes, pitch, timing, velocity, and any other specific control parameter.

MIDI messages are structured with an 8-bit status byte followed by one or two data bytes, where the status byte indicates the type of message being sent [14]. These messages are classified into Channel Messages, which are specific to individual channels, and System Messages, which are not channel-specific and are received by all devices. The first byte carries the status message and is used to send musical performance information, such as Note On, Note Off, Polyphonic Key Pressure, Channel Pressure, Pitch Bend Change, Program Change, and Control Change messages [14]. The second byte carries the information regarding the note number such as C4, C#4, D4 etc. of the scale translated to numbers. For example, C4 on an electric piano is stored as '72' and D4 is stored as 74. The third byte tells us regarding the velocity of the note pressed, ranging from 0 – 127. For example, when a user presses a note C in the 4$^{th}$ octave on a keyboard connected to a computer, the keyboard would first recognize the note pressed and the intensity with which it was pressed and send a message containing the action, the note number, and the intensity. The message would read 'Note on, C4, 100' and this is translated through the MIDI protocol as three digits with '144' being the code for 'Note on', '72' being 'C4' and the third digit being the velocity number. When the finger on the note is removed, it sends another signal this time reading '128, 72, 0' which reads 'Note off, C4, Velocity:0'. For Control Change messages, for example of CC 1 message, it would read '146, 1, 100' where '146' is code for CC messages, '1' is the Controller Number and '100' is the intensity. A breakdown of this translation is provided in Figure 4.

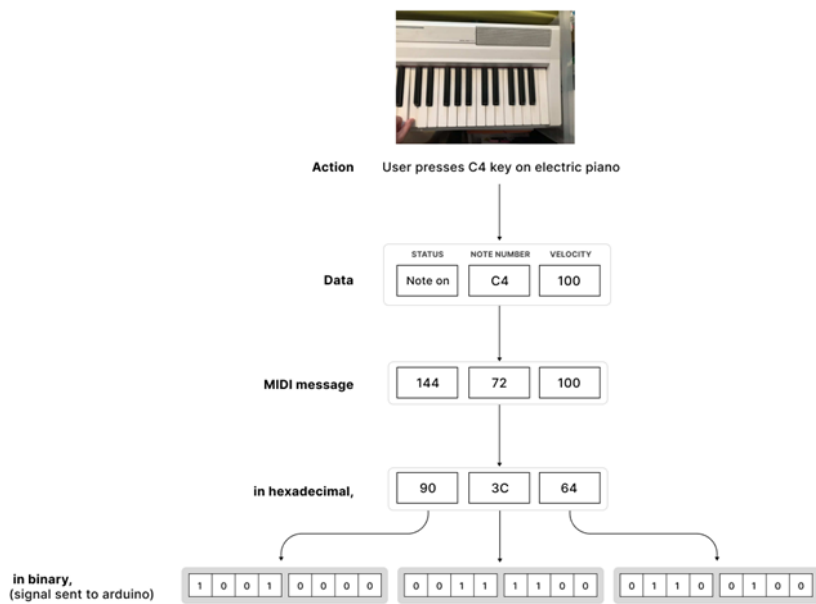

**Fig. 4.** Working Method of MIDI protocol

In this strategy, signals are transmitted from the computer using a Digital Audio Workstation (DAW), where a MIDI composition can be created and edited, directly to the Arduino. The code in Arduino reads the three bytes of midi messages received and maps the note number and control value to the corresponding haptic actuator. To validate the functionality of the MIDI protocol, a preliminary test was conducted using Supercollider [15]. A sample project was implemented to generate MIDI Continuous Controller (CC) messages that interacted with a graphical user interface (GUI), simulating drum playback. The MIDI output from this test was then linked to a virtual MIDI port through loopMIDI [16]. Given that communication between the Arduino and PC occurred through a USB Serial port, integration with the MIDI protocol was facilitated by the utilization of a MIDI to USB Serial bridge program named Hairless MIDI [17]. This workflow is presented in Figure 5.

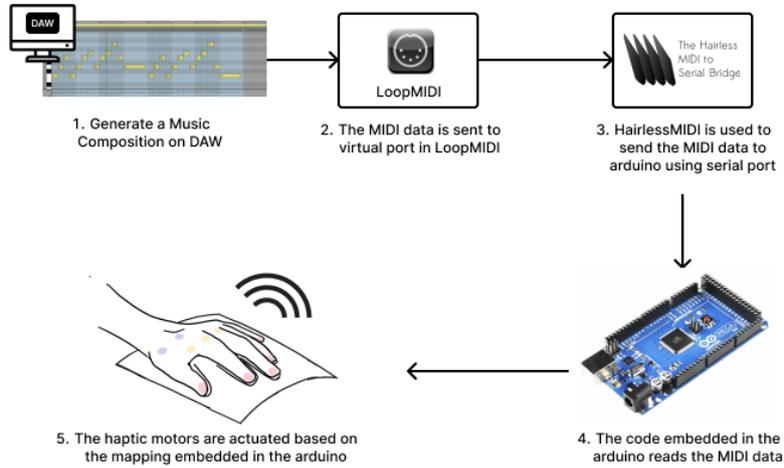

**Fig. 5.** Workflow of the System

In the code, which is burnt to the Arduino Mega, we employ the use of both MIDI messages and CC messages as they both offer unique ability to control the actuators. MIDI messages are much easier and more direct to create. By simply playing a melody or importing the MIDI file, one can instantly play and edit the music on the haptic music player. However, it limits the intensity of the actuator to a constant value, which is set based on the velocity of the note pressed. Hence, we are not able to manipulate the haptic signature of the melody being played, restricting the timbre aspect of music. The CC messages, however, provide much more flexibility in manipulating the haptic signature, by being able to create custom graphs on how the intensity should vary temporally. However, its limitations arise with the tedious process of producing music using these messages.

## 4    Music to Haptics Mapping

### 4.1    Background of Haptic Music Players

The method of communicating music through audio already involves a vibrational aspect of a speaker's diaphragm. The signals of the audio file, when directly mapped to the haptic actuators reproduce the loudness in a compressed frequency range limited to the range of the actuator. However, this method does not effectively translate the other elements of music such as melody, pitch, rhythm, and timbre. Hence, a more comprehensive Music to Haptic mapping strategy was required.

Looking at some of the haptic music players by current researchers, we can find most of them started by offering a full body haptic experience, by using frequency analysis on the audio and filtering frequency bands and spatially mapping it to the placement of actuators. The Emoti-chair [18] used this strategy by mapping the lowest frequencies at the bottom, under the seat and higher frequencies travelling up the spine of the user. In more recent wearable haptic music players, such as Feel-Music [4] and Mood Glove [5] the 8 notes of the musical octave (C, D, E, F, G, A, B, C) were each assigned to an individual actuator and arranged spatially on the forearm for Feel Music, and on the dorsal side of the palm. This method of focusing on translating the melody spatially, has reportedly both enhanced the perception of music when audio and tactile feedback is combined, and users were able to recognize the audio counterpart through the haptic song. [4][5]

### 4.2     Haptic Mapping of Our Device

Most of the music produced currently is primarily made up of two layers, the Harmonics, and the Percussions. To understand the mapping of audio to haptics, we must first understand each of them individually to better translate the haptic version of the music. Harmonics in music refer to the higher-frequency components that accompany the fundamental frequency of a musical note or sound, contributing to the timbre and character of the sound [19]. Examples of harmonics in music include string instruments like guitars and violins, flute harmonics, and singing harmonics. Percussive sounds in music are characterized by a clash, knock, or transient attack phase, often represented by vertical lines in a spectrogram representation [20]. Examples of percussive instruments include drums, castanets, xylophones, and marimbas, which produce sound through striking or shaking, resulting in a wide range of tonal colors and textures [20][19].

### 4.3     Haptic Mapping of Harmonics

There are two ways to go about this. First, to directly map the frequency of the note to the frequency of the motor. Thereby varying the intensity of the actuator to communicate pitch. Or the second, to spatially arrange notes on the hand and assign an individual actuator. The second method frees up the intensity parameter of the actuator to communicate the timbre of the sound. The challenge is to now arrange the twelve notes to be spatially perceptible. In the harmonics layer we have the main leading melody underlying chords.

Based on the touch sensitivity distribution discussed in Section 3.2, the main leading melody is best arranged spatially under the five fingertips due to its higher sensitivity to pick up timbre of the instrument as well. Since there are twelve notes but only five actuator regions, we employed the use of a tactile illusion known as "cutaneous rabbit illusion" [21] to create the feeling of the pitch coming from an imaginary circular disc under the hand. The tactile illusion of perceiving a sense of touch between two fingers when they are actuated alternatively is known as the "cutaneous rabbit illusion"[21].

This illusion occurs when a sequence of taps at two separated skin locations on the fingers results in the perception that the region between the fingers are also tapped, creating the sensation of a "rabbit" hopping between the fingers [21]. These exteroceptive signals can be used to expand the range of perceiving melodies without losing space.

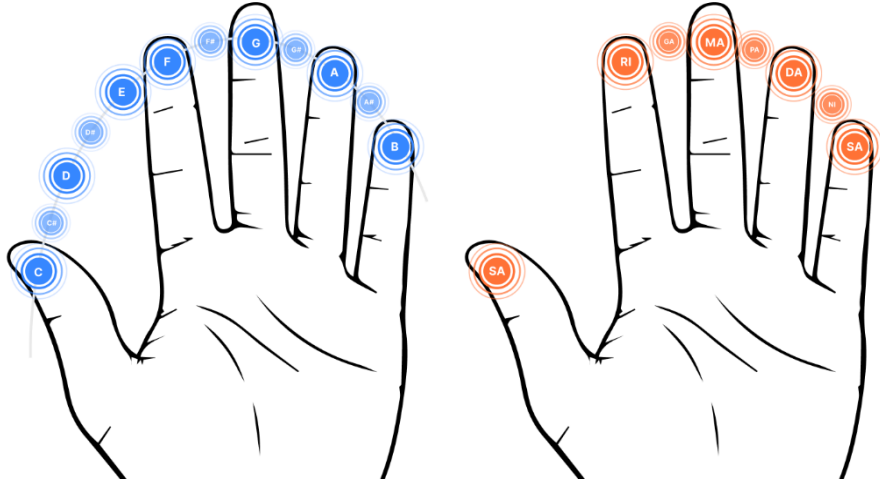

**Fig. 6.** Spatial Mapping of notes onto fingers in Western Octave (left) and Indian Swaras (right)

Due to the low tactile spatial acuity present in the metacarpophalangeal, thenar and hypothenar regions, spatial arrangement of notes is seen to be less effective. Hence, here the intensity of the notes is mapped to the three actuators in the metacarpophalangeal region. This way provides a more holistic translation of the melodies without overshadowing the main leading melody. Songs with an underlying baseline below the chorus section can take advantage of the thenar and hypothenar region and map the pitch to the intensity of the motors to create a deeper impression of the song.

### 4.4 Haptic Mapping for Piano

Taking the example of a famous classical piano piece, Ludwig Van Beethoven's Für Elise, we can understand the different variations present in the performance and how these variations are translated to the desk-mounted haptic player. The piece is widely acknowledged since the first eight bars can be adapted into a limited "sonic palette" better than most other classical works.[22]

The sheet music of Für Elise features a beautiful, singing melody in the right hand, which is supported by flowing, arpeggiated chords in the left hand. The melody primarily occupies the middle and upper octaves, while the accompanying chords span a wide range from the low to middle octaves, creating a rich, harmonically complex texture.

The main melody of Für Elise is played by the right hand. It starts in the middle octave, with the 4th finger on E and the 3rd finger on D#. The melody consists of a repeating pattern of E-D#-E-D#-E, B-D-C-A. These notes are spatially mapped onto the tips actuators of the device as demonstrated in Figure 4.2. The melody often moves between the middle and upper octaves. This can be translated by using specific ranges of intensity of the actuators based on the octave which is being played. For example, if the melody is being played in the middle octave, the actuators can only operate in the range of 50 Hz to 100 Hz (or an intensity of 85 to 170), and when played in the upper octave can employ the range 100 Hz to 150 Hz (or an intensity of 170 to 255).

The left-hand plays arpeggiated (broken) chords that accompany the melody. It starts in the lower octave. The left hand plays a repeating pattern of A-E-A, moving up to the middle octave. These arpeggiated chords provide harmonic support and a flowing, rippling effect underneath the melody. The left-hand chord progressions move between A minor, C major, and other related keys, creating a sense of harmonic movement. The left-hand chords often span a wide range, from the low A to the middle A or higher. These notes are mapped to the middle actuators on the metacarpophalangeal region with pitch directly correlated to the frequency of the actuators. Varying from 50 Hz to 150 Hz covering two octaves (24 notes) with each semitone at intervals of 4.16 Hz.

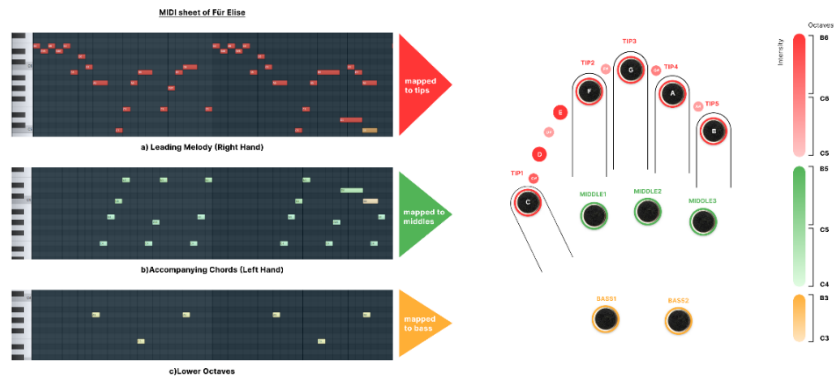

**Fig. 7.** Haptic Mapping of Piano for Für Elise

### 4.5   Haptic Mapping of Percussions

Percussion instruments have a strong connection to the sense of touch related to haptics due to the intricate relationship between touch and sound production in these instruments. The physical interaction with percussion instruments involves a direct tactile experience that influences the quality and characteristics of the sound produced. Touch receptors in the skin are much better at reading sharp, high-impact signals like the sound percussion instruments produce [3]. The focus would now go into the distribution of the diverse types of percussion sounds on the palm of the hand since the signature of the signal is same in audio and haptics.

Taking examples of the sounds of drums we can discuss the haptic mapping of these sounds onto the palm. As decided previously, the higher frequency and sharper sounds such as snare drum, hi-hats, crash cymbals and ride cymbals are better attributed to the fingers (tips actuators). Whereas the low frequency sounds such as bass drum are attributed to the thenar and hypothenar region (bass actuators). The Toms, which fall in the middle frequencies are attributed to the metacarpophalangeal region (middle actuators) to add flavor to the sound. The snare and bass drums are the most important percussion instruments in a drum kit due to their essential roles in establishing the rhythmic foundation, dynamic range, and overall timbre of the music [23]. The snare drum provides the rhythmic accents and bright, high-pitched sounds that drive the beat, while the bass drum sets the pulse and low-end foundation, making them indispensable for a wide range of musical genres from classical to rock. To transpose this nature of these drums, the snare drum and kick drum are also attributed to the middle actuators to generate stronger haptic impact, as their waveforms do in audio format.

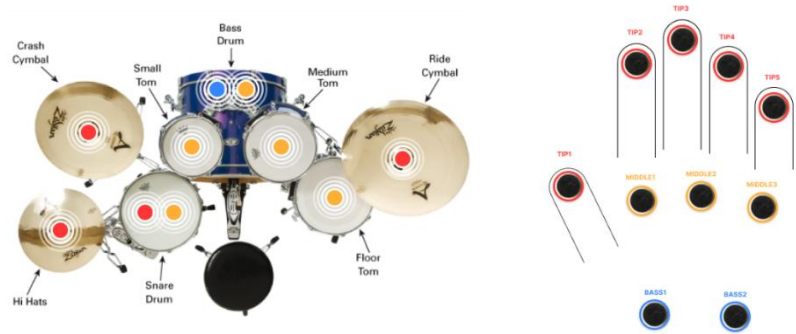

**Fig. 8.** Instrument to Actuator mapping of an acoustic drum kit

## 5 User Study

A preliminary study was conducted to test the efficacy of translating music to haptics, and to understand musical haptics by evaluating the functionality, usability, and user experience of haptic feedback devices. The data from the participants was obtained using a Likert scale to measure their responses.

### 5.1 Experiment Setup

The participant was instructed to listen to a short piano snippet taken from John Thompson's Modern Course for Piano: first grade in audio format. The test was designed to find whether the participant could identify the song correctly when the music snippet was played through the haptic ERM motors. Figure 5.1 shows the experimental setup containing the DAW and our haptic interface for testing.

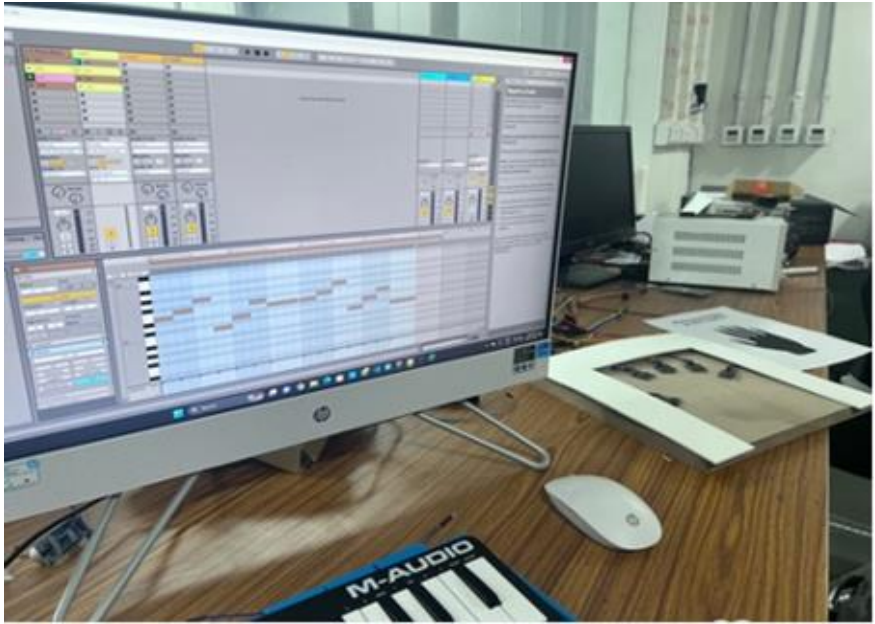

**Fig. 9.** Experimental Setup

### 5.2     Participants

A set of 8 participants from in and around the campus with an average age of 23.5 years (7 male and 1 female) participated in the experiment. 3 out of them were musically trained, of which 2 of them were able to play piano. The rest of the remaining 5 participants did not have any music training. Prior consent and approvals necessary for the experiment were taken.

### 5.3     Material

The experiments were performed on a Windows 11 PC with a 3.7 GHz Intel Core-i3 12th gen processor and 8.0 GB RAM. The vibrotactile feedback is generated through five Eccentric Rotating Mass (ERM) Coin Vibrators (5V, 60mA, 9000RPM). The communication between PC and ERM is done via an Arduino Mega using a Serial USB port. The music is composed and played on Ableton and sent the MIDI out through a virtual MIDI port using LoopMIDI. Hairless MIDI is then used as a bridge between MIDI to Serial to send the data to said Arduino Mega. The system is mounted on a wooden plank for comfortable resting of participant's hands.

### 5.4     Design

For this research study, we selected three musical snippets from John Thompson's Modern Course for Piano: first grade [24]. The three snippets chosen were Music Land (Song 1), Patterns (Song 2) and Traffic Cop (Song 3). The treble clef notes of these snippets were transcribed into a digital audio workstation (DAW) as MIDI data for analysis (Figure 5.2). The haptic mapping of each ERM motor was employed based on the right-hand finger positions outlined in John Thompson's instructional material. If a note is played in MIDI, the motor-in-contact with the finger corresponding to the finger number (w.r.t John Thompson) will be triggered. Given the variability in finger positioning among musicians, influenced by individual comfort and dexterity levels, we aimed to establish a standardized mapping function applicable to all participants. We chose to refer to John Thompson's instructional material due to its inclusion of simple and recognizable melodies with corresponding finger positions tailored for beginner pianists.

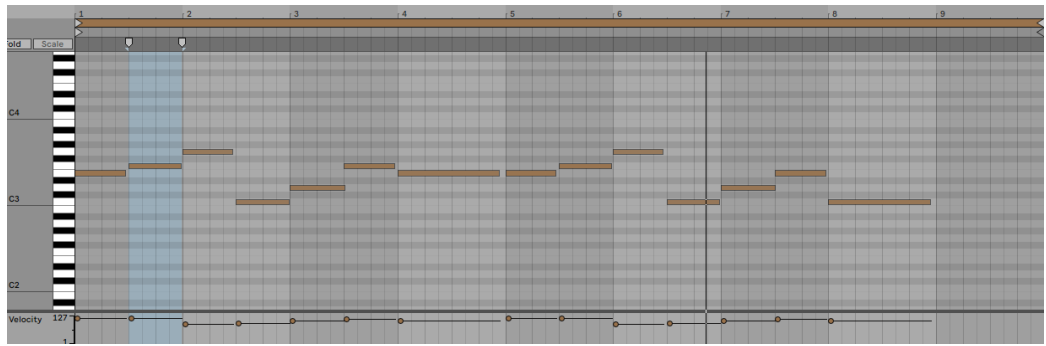

(a)

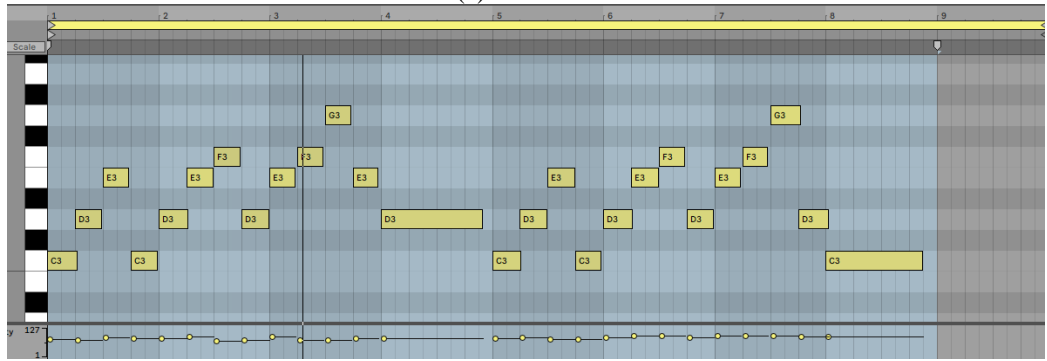

(b)

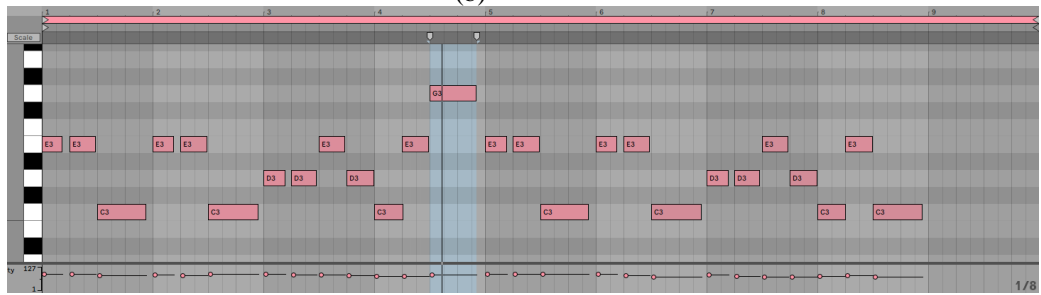

(c)

**Fig. 10.** MIDI Sheet of (a) Song 1, (b) Song 2 and (c) Song 3

### 5.5   Procedure

Participants were first introduced to the three songs through auditory exposure until they memorized the melodies. Subsequently, participants were required to accurately identify the song labels by listening to the audio, thus mitigating the likelihood of recall failure. Following this auditory familiarization phase, participants were introduced to haptic renditions of the melodies

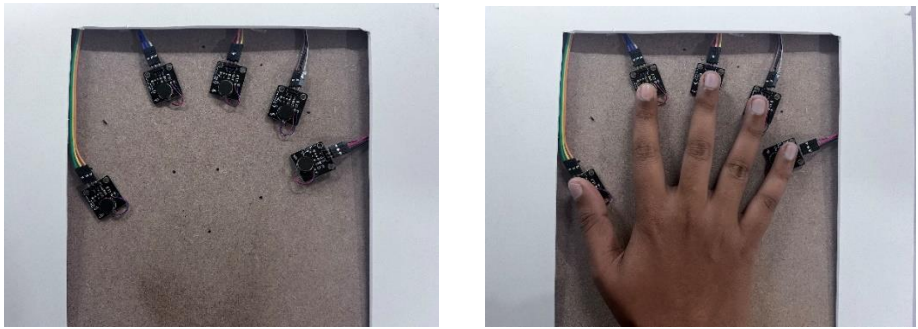

**Fig. 11.** Haptic Interface (left) and Fingertips on Interface (right)

Following each trial, participants were prompted to recognize the song label when its haptic version was played in our interface as shown in Figure 11. This iterative process ensured the participants' ability to recognize song labels with minimal errors. Upon achieving proficiency in the practice session, participants were tasked with ten additional trials aimed at identifying the song labels corresponding to the perceived haptic versions. Responses, categorized as correct or incorrect labels, were recorded in a spreadsheet. Subsequent analysis focused on assessing the accuracy of participants' identification of the correct song label. Additionally, participants were asked to rate their confidence levels in label identification.

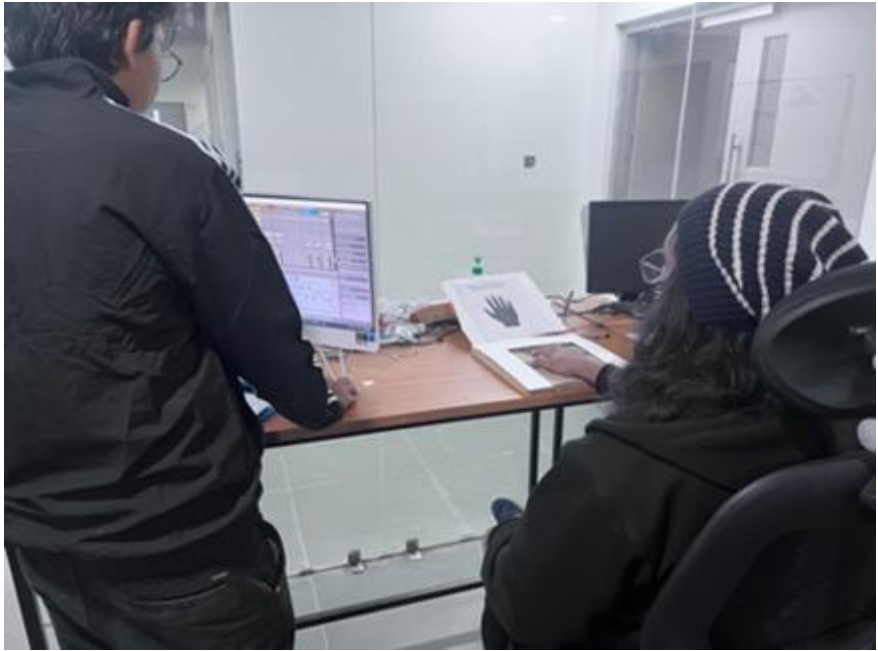

**Fig. 12.** Participant undertaking trial

## 6 Results

In this section, we discuss the results in terms of the accuracy and confidence of the partici-pants, corresponding to their musical training. We also discuss the recall and precision data of each song based on the responses. The results are as follows:

### 6.1 Accuracy of recognition

The participants were given 10 trials each, playing the three songs through haptic systems in a random sequence. Their answers were recorded to see how many were correctly identified. We also noted the participants' musical training. In Figure 7, we can infer that 5 of the participants were 100% accurate in recognizing the songs. We also found that 3 musically untrained participants made few mistakes and had an accuracy of 80% compared to the 100% accuracy of musically trained participants.

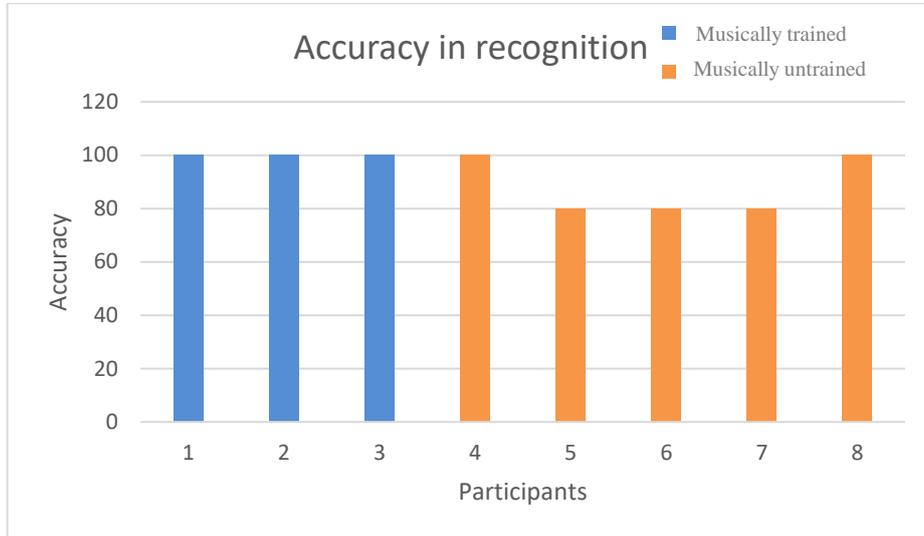

**Fig. 13.** Accuracy of recognizing songs

### 6.2   Confidence in recognition

After each trial, the participants were also prompted to provide a self-assessment of their confidence level regarding the accuracy of their song label identification. The confidence ratings were collected on a Likert scale, ranging from 1 to 10, with participants indicating their perceived confidence in their ability to correctly identify the given song, with 10 being most confident and 1 being least confident. From Figure 8, we see that the musically trained participants were more confident than the musically untrained participants.

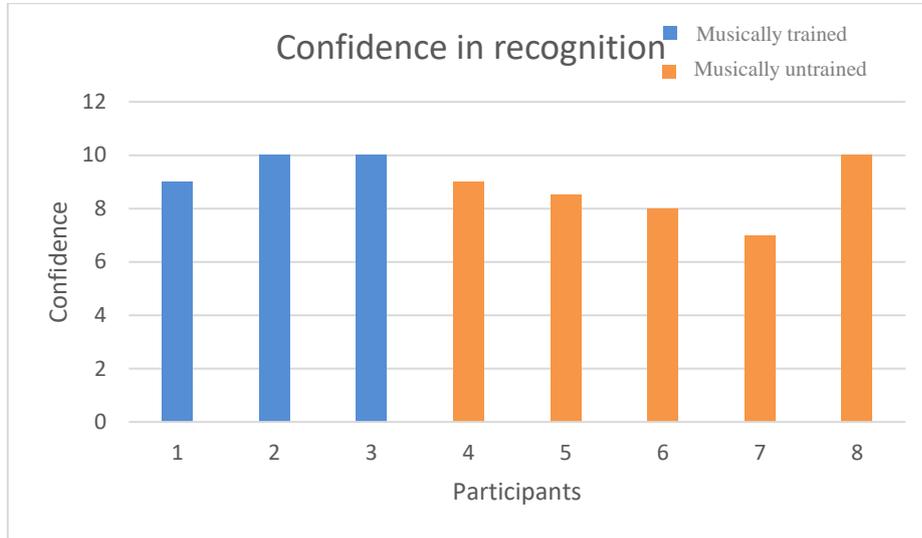

**Fig. 14.** Precision of recognizing songs

### 6.3   Precision and Recall

We calculated the precision and recall for each song based on the recognition results by the participants. We calculate the precision and recall for a song (song 1) as follows.

$$\text{Precision} = \frac{\text{Total number of correct ones}}{\text{Total number of detected ones}}$$

$$\text{Recall} = \frac{\text{Total number of correct ones}}{\text{Total number of given ones}}$$

Similarly, we calculated precision and recall for songs 2 and 3 and reported the values in Table 1

**Table 1.** Precision and recall for recognizing each song

| Song Label | Precision | Recall |
|---|---|---|
| Song 1 | 0.94 | 0.94 |
| Song 2 | 0.91 | 0.83 |
| Song 3 | 0.92 | 1.00 |

## 7      Discussion

From the results we can infer that the difference in accuracy of recognition between musically trained and untrained participants was less (20%). In most of the trials, the participants were able to correctly recognize the song from the haptic version played on the interface. When the participants were asked how they felt about the interface and their experience in identifying the songs, they said they were able to make out the differences in the pattern of haptics for each song and identify. The musically trained participants reported that they could identify the rhythm corresponding to each song through haptics and recognize the songs accordingly. This clearly indicates our interface could translate certain musical elements via characteristics inherent to haptics.
The confidence scores show that the musically trained participants were more confident in recognizing the songs than the musically untrained. This might be due to the influence of their musical skills and aptitude in perceiving rhythm compared to the untrained.

The precision and recall values indicate that song 2 has the least recall value. This might be due to the confusion between songs 2 and 3 as reported by some of the participants. Overall, the participants liked the interface and were able to recognize the songs and musical patterns through haptics without audio through our interface.

## 8      Future Scope

Further investigation into complex musical structures and mapping with different textures for more resolution and enhanced experience would be essential in upcoming studies. Investigation of the interface for people with hearing impairment and see if it has a potential in sensory substitution. HMPs have the potential to enrich music listening activities and provide opportunities for the hearing impaired to engage with music in a new way. However, the development of affordable and accessible haptic devices with adaptive and individually tuned feedback is a challenge that needs to be addressed. Further into the research, we can try to incorporate wider variety of vibrotactile motors to translate diverse textures of sounds, conducting a study on the perception of texture on sound. Long term studies on the addition of haptic information along with visual and auditory information in the process of learning a new instrument can be conducted as well. With studies showing students with haptic training to be more accurate when it's incorporated in teaching piano [25], the system can also be used to incorporate various western and traditional instruments, primarily percussion instruments such as Tabla, Mridangam.

Additionally, the desk-mounted haptic system can also be used in adjacent areas of research as well, including medical and mental-health applications. A study found that mindfulness exercises in VR increase state mindfulness and positive emotions. Passive haptic feedback, such as feeling grass under one's hands, became more important during the mindfulness task, increasing presence and focus. Haptics-assisted meditation (HAM) incorporates force control and meditation practice. During HAM, participants maintain awareness on their respiration while adjusting bimanual fingertip pressures to keep synchronized with their breathing. This paradigm requires somatosensory attention and helps novices meditate starting with the body to gradually gain essential meditation skills. A study showed that HAM reduced mind-wandering compared to classic breath-counting meditation [27].

## 9 Conclusion

We developed a desk-mounted haptic interface capable of translating musical pieces from MIDI to haptics, thereby allowing the song to be experienced in both audio and haptic formats. A user study was conducted to evaluate the efficacy of our interface, specifically focusing on the accuracy of song recognition through haptics in the absence of audio. The results indicated promising outcomes in terms of accuracy, precision, and recall for song recognition via haptics. Participants reported positive feedback regarding their experience with the interface, although there remains significant scope for enhancement.

This project facilitated my exploration and understanding of the field of Human-Computer Interaction (HCI), introducing me to the design and creation of electronic devices using Arduino, which involves the integration of coding, sensors, and actuators. The desk-mounted haptic system we prototyped demonstrates versatility and adaptability, making it suitable for various studies and applications. Additionally, it holds potential as an entertainment device with diverse use cases. Future research will focus on exploring the complexities of musical mapping to haptics and examining the utility of our interface for individuals with hearing impairments, in the learning of musical instruments and in the mental health & meditation sector.

**Disclosure of Interests.** It is now necessary to declare any competing interests or to specifically state that the authors have no competing interests. Please place the statement with a third level heading in 9-point font size beneath the (optional) acknowledgments[1], for example: The authors have no competing interests to declare that are relevant to the content of this article. Or: Author A has received research grants from Company W. Author B has received a speaker honorarium from Company X and owns stock in Company Y. Author C is a member of committee Z.

**References**

1. Boudry, C., Chartron, G.: Availability of digital object identifiers in publications archived by PubMed. Scientometrics 1-26 (2017). https://hal.science/hal-04029009/document
2. Papetti, S., Saitis, C. (eds.): Musical Haptics. Springer International Publishing, Cham (2018). https://www.loc.gov/item/2019753348/
3. Remache-Vinueza, B., Trujillo-León, A., Zapata, M., Sarmiento-Ortiz, F., Vidal-Verdú, F.: Audio-Tactile Rendering: A Review on Technology and Methods to Convey Musical Information through the Sense of Touch. Sensors (Basel) 21(19), 6575 (2021). https://api.semanticscholar.org/CorpusID:238745704
4. Haynes, A., Lawry, J., Kent, C., Rossiter, J. M.: FeelMusic: Enriching Our Emotive Experience of Music through Audio-Tactile Mappings. Multimodal Technologies and Interaction 5(6), Article 29 (2021). https://doi.org/10.3390/mti5060029
5. Mazzoni, A., Bryan-Kinns, N.: Mood Glove: A haptic wearable prototype system to enhance mood music in film. Entertainment Computing 17, 9-17 (2016). https://doi.org/10.1016/j.entcom.2016.06.002
6. Lane-Smith, J., Chow, D., Ajami, S., Cooperstock, J.: The Hapstrument: A Bimanual Haptic Interface for Musical Expression. In: International Conference on New Interfaces for Musical Expression (2023). https://doi.org/10.5281/zenodo.11189284
7. Wegen, M., Herder, J., Adelsberger, R., Pastore-Wapp, M., van Wegen, E., Bohlhalter, S., Nef, T., Krack, P., Vanbellingen, T.: An Overview of Wearable Haptic Technologies and Their Performance in Virtual Object Exploration. Sensors 23, 1563 (2023). https://doi.org/10.3390/s23031563
8. Remache Vinueza, B., Trujillo-León, A., Zapata, M., Sarmiento Ortiz, F., Vidal-Verdú, F.: Audio-Tactile Rendering: A Review on Technology and Methods to Convey Musical Information through the Sense of Touch. Sensors 21, 6575 (2021). https://doi.org/10.3390/s21196575
9. Hu, X., Kando, N.: Task complexity and difficulty in music information retrieval. Journal of the Association for Information Science and Technology 68(7), 1711–1723 (2017). https://doi.org/10.1002/asi.23803
10. Abraira, V. E., Ginty, D. D.: The Sensory Neurons of Touch. Neuron 79(4), 618-639 (2013). https://doi.org/10.1016/j.neuron.2013.07.051
11. Home Science Tools: Sense of Touch, Skin Receptors, Skin Sensations, Somatosensory System. Home Science Tools Homeschool Hub (2024). https://learning-center.homesciencetools.com/article/skin-touch/
12. Stark, B., Carlstedt, T., Hallin, R. G., Risling, M.: Distribution of Human Pacinian Corpuscles in the Hand: A cadaver study. Journal of Hand Surgery 23(3), 370-372 (1998). https://doi.org/10.1016/S0266-7681(98)80060-0
13. Chakrabarti, D.: Indian Anthropometric Dimensions for Ergonomic Design Practice. National Institute of Design (1997)
14. Sunderland, D.: MIDI Messages (2024). https://users.cs.cf.ac.uk/dave/Multimedia/node158.html
15. McCartney, J.: Rethinking the computer music language: Super collider. Computer Music Journal 26(4), 61-68 (2002)
16. Ericcson, T.: LoopMIDI. (2010). https://www.tobias-erichsen.de/software/loopmidi.html (accessed Feb. 8, 2024)
17. Gratton, A.: HairlessMIDI. (2011). https://projectgus.github.io/hairless-midiserial/
18. Karam, M., Branje, C., Nespoli, G., Thompson, N., Russo, F., Fels, D.: The emoti-chair: An interactive tactile music exhibit. In: Proceedings of the 28th International Conference on

Human Factors in Computing Systems, pp. 3069-3074 (2010). https://doi.org/10.1145/1753846.1753919
19. EduInput: 10 Example of Harmonic in Music (2023). https://eduinput.com/example-of-harmonic-in-music/
20. LiberTexts: Harmonic Percussion Instruments (2021). https://chem.libretexts.org/Bookshelves/Physical_and_Theoretical_Chemistry_Textbook_Maps/Supplemental_Modules_(Physical_and_Theoretical_Chemistry)/Spectroscopy/Vibrational_Spectroscopy/Harmonic_Percussion_Instruments
21. Geldard, F. A., Sherrick, C. E.: The cutaneous "rabbit": A perceptual illusion. Science 178(4057), 178-179 (1972)
22. Wikipedia: Für Elise. (2024). https://en.wikipedia.org/wiki/F%C3%BCr_Elise
23. Pratt, G.: The dynamics of drumming. Schott, London (1998)
24. Thompson, J.: John Thompson's Modern Course for the Piano-Second Grade (Book Only): Second Grade. Hal Leonard Corporation (2005)
25. Tom, A., Singh, A., Daigle, M., Marandola, F., Wanderley, M. M.: Haptic Tutor - A haptics-based music education tool for beginners. In: HAID 2020 - International Workshop on Haptic and Audio Interaction Design, Montreal / Virtual, Canada (2020). ffhal-02901205f
26. Zheng, Y. L., Wang, D. X., Zhang, Y. R., Tang, Y. Y.: Enhancing Attention by Synchronizing Respiration and Fingertip Pressure: A Pilot Study Using Functional Near-Infrared Spectroscopy. Frontiers in Neuroscience 13, 1209 (2019). https://doi.org/10.3389/fnins.2019.01209